\documentclass[onecolumn,amsmath,amssymb,11pt,superscriptaddress,nofootinbib]{revtex4}
\pdfoutput=1
\usepackage[bookmarks,linktocpage, colorlinks=true, plainpages = false, citecolor = blue,  linkcolor=blue, urlcolor = blue,
filecolor = blue]{hyperref}

\usepackage{graphicx}
\usepackage{dcolumn}
\usepackage{bm}
\usepackage{braket}
\usepackage{verbatim} 
\usepackage[]{inputenc}
\usepackage{cancel}
\usepackage[usenames,dvipsnames]{color}
\usepackage{natbib}

\input epsf
\def\gsim{ \lower .75ex \hbox{$\sim$} \llap{\raise .27ex \hbox{$>$}} }
\def\lsim{ \lower .75ex \hbox{$\sim$} \llap{\raise .27ex \hbox{$<$}} }

\newcommand{\be}{\begin{equation}}
\newcommand{\ee}{\end{equation}}
\newcommand{\ba}{\begin{aligned}}
\newcommand{\ea}{\end{aligned}}
\newcommand{\bea}{\begin{eqnarray}}
\newcommand{\eea}{\end{eqnarray}}
\renewcommand{\d}{\mathrm{d}}

\def\g{\gamma}

\def\de{\delta}
\def\z{\zeta}

\def\ep{\epsilon}
\def\f{\phi}
\def\fd{\dot{\phi}}
\def\fdd{\ddot{\phi}}

\def\F{\Phi}

\def\m{\mu}
\def\n{\nu}

\def\x{\xi}
\def\z{\zeta}
\def\p{\partial}

\newcommand{\cH}{\mathcal{H}}

\newcommand{\h}{\frac{1}{2}}

\begin{document}

	\begin{titlepage}
		
		\title{Conflation: a new type of accelerated expansion}
		
		\author{Angelika Fertig}
		\email[]{angelika.fertig@aei.mpg.de}
		\author{Jean-Luc Lehners}
		\email[]{jlehners@aei.mpg.de}
		\author{Enno Mallwitz}
		\email[]{enno.mallwitz@aei.mpg.de}

		\affiliation{Max--Planck--Institute for Gravitational Physics (Albert--Einstein--Institute)\\ Am M\"{u}hlenberg 1, 14476 Potsdam-Golm, Germany}
		
		\begin{abstract}
\vspace{1cm}
\noindent 
In the framework of scalar-tensor theories of gravity, we construct a new kind of cosmological model that conflates inflation and ekpyrosis. During a phase of conflation, the universe undergoes accelerated expansion, but with crucial differences compared to ordinary inflation. In particular, the potential energy is negative, which is of interest for supergravity and string theory where both negative potentials and the required scalar-tensor couplings are rather natural. A distinguishing feature of the model is that, for a large parameter range, it does not significantly amplify adiabatic scalar and tensor fluctuations, and in particular does not lead to eternal inflation and the associated infinities. We also show how density fluctuations in accord with current observations may be generated by adding a second scalar field to the model. Conflation may be viewed as complementary to the recently proposed anamorphic universe of Ijjas and Steinhardt. 
\end{abstract}
\maketitle

\end{titlepage}

\tableofcontents
\section{Introduction}

Inflation \cite{Starobinsky:1980te,Kazanas:1980tx,Sato:1980yn,Guth:1980zm,Linde:1981mu,Albrecht:1982wi} and ekpyrosis \cite{Khoury:2001wf} share a number of features: they are the only dynamical mechanisms known to smoothen the universe's curvature (both the homogeneous part and the anisotropies) \cite{Guth:1980zm,Erickson:2003zm}. They can also amplify scalar quantum fluctuations into classical curvature perturbations which may form the seeds for all the large-scale structure in the universe today \cite{Mukhanov:1981xt,Lehners:2007ac}. Moreover, they can explain how space and time became classical in the first place \cite{Lehners:2015sia}. With a number of assumptions, in both frameworks models can be constructed that agree well with current cosmological observations, see e.g. \cite{Martin:2013nzq,Lehners:2013cka}. But in other ways, the two models are really quite different: inflation corresponds to accelerated expansion and requires a significant negative pressure, while ekpyrosis corresponds to slow contraction in the presence of a large positive pressure. Inflation typically leads to eternal inflation giving rise to the measure problem \cite{Ijjas:2013vea, Guth:2013sya}, 
while ekpyrosis requires a null energy violating (or a classically singular) bounce into the expanding phase of the universe \cite{Lehners:2011kr}.

In the present paper, we will present a new cosmological model that combines features of both inflation and ekpyrosis. This is in the same spirit as the recently proposed ``anamorphic'' universe of Ijjas and Steinhardt \cite{Ijjas:2015zma}, the distinction being that we are combining different elements of these models. We will work in the framework of scalar-tensor theories of gravity. By making use of a field redefinition (more precisely a conformal transformation of the metric), we transform an ekpyrotic contracting model into a phase of accelerated expansion. Moreover, we are specifically interested in the situation where matter degrees of freedom couple to the new (Jordan frame) metric, so that observers made of this matter will measure the universe to be expanding. Conflation is reminiscent of inflation in the sense that the background expands in an accelerated fashion. This then immediately implies that the homogeneous spatial curvature and anisotropies are diluted, thus providing a solution to the flatness problem. However, other features of the model are inherited from the ekpyrotic starting point of our construction: for instance, the model assumes a negative potential. This might have implications for supergravity and string theory, where negative potentials arise very naturally and where it is in fact hard to construct reliable standard inflationary models with positive potentials \cite{Dasgupta:2014pma}. Also, for a large parameter range conflation does not significantly amplify adiabatic curvature perturbations (nor tensor perturbations). Hence eternal inflation, which relies on the amplification of large, but rare, quantum fluctuations, does not occur. This has the important consequence that the multiverse problem is avoided. As we will show, one can however obtain nearly scale-invariant curvature perturbations by considering an entropic mechanism analogous to the one used in ekpyrotic models \cite{Li:2013hga,Qiu:2013eoa,Fertig:2013kwa,Ijjas:2014fja,Levy:2015awa}. This allows the construction of specific examples of a conflationary phase in agreement with current cosmological observations.    

For related studies starting from an inflationary phase and transforming that one into other frames, see \cite{Piao:2011bz,Qiu:2012ia,Wetterich:2014eaa,Li:2014qwa,Domenech:2015qoa}, 
while \cite{Boisseau:2015hqa} studies a related scanario of inflation preceeded by a bounce. %
In the language of the anamorphic universe \cite{Ijjas:2015zma}, we are looking at the situation where $\Theta_m>0$ and $\Theta_{Pl}<0,$ while Ijjas and Steinhardt consider $\Theta_m < 0$ and $\Theta_{Pl}>0$ (note that inflation corresponds to $\Theta_m > 0$ and $\Theta_{Pl}>0$ and ekpyrosis to $\Theta_m < 0$ and $\Theta_{Pl}<0$).

\section{Ekpyrotic Phase in Einstein Frame}\label{sec:EkpE}
We start by reviewing the basics of ekpyrotic cosmology \cite{Khoury:2001wf,Lehners:2008vx}. During an ekpyrotic phase the universe undergoes slow contraction with high pressure $p.$ The equation of state is assumed to be large, $w=p/\rho > 1,$ where $\rho$ denotes the energy density of the universe. Under these circumstances both homogeneous curvature and curvature anisotropies are suppressed, and consequently the flatness problem can be resolved if this phase lasts long enough. The ekpyrotic phase can be modelled by a scalar field with a steep and negative potential, with action (in natural units $8 \pi G=M_{Pl}^{-2}=1$)
\be \label{eq:Eaction}
S = \int \d^4 x \sqrt{-g} \left[\frac{R}{2} - \h g^{\mu \nu }\p_{\mu} \f \p_{\nu} \f - V(\f) \right]\,,
\ee
where a typical ekpyrotic potential is provided by a negative exponential,
\be
V(\f) = - V_0 e^{- c \f}\,.
\ee 
We consider a flat Friedmann-Lema\^{i}tre-Robertson-Walker (FLRW) universe, with metric $\d s^2 = - \d t^2+ a(t)^2 \de_{ij} \d x^i \d x^j$, where $a(t)$ is the scale factor and with $\dot{} \equiv \d /\d t$.
The equation of motion for the scalar field is then obtained by varying the action w.r.t. the scalar field $\f$
\be
\fdd + 3 H \fd + V_{,\f}=0,
\ee
and it admits the (attractor) scaling solution \cite{Khoury:2001wf}
\be \label{eq:phi}
a(t)=a_0 \left( \frac{t}{t_0} \right)^{^1/_\ep}, \;\;\;\; \f = \sqrt{\frac{2}{\ep}} \ln{\left(\frac{t}{t_0}\right)}, \;\;\;\;
\text{where} \;\; t_0 = - \sqrt{\frac{\ep-3}{V_0 \ep^2}}
\;\;\;\;
\text{and} \;\; c= \sqrt{2 \ep}.
\ee
The coordinate time $t$ is negative and runs from large negative values towards small negative values. The fast roll parameter $\epsilon=\frac{\dot \f^2}{2H^2}$ is directly related to the equation of state 
$w=\frac{2}{3}\epsilon-1,$ while the condition that an ekpyrotic phase has to satisfy, $w>1$, is equivalent to $\epsilon>3$.

\section{Conflation}\label{sec:transfJ}
The above model was constructed in the standard Einstein frame where the scalar field is minimally coupled to gravity. In the following we perform a conformal transformation to the so-called Jordan frame, where the scalar field is now non-minimally coupled to gravity.

\subsection{Jordan frame action}
A general transformation to Jordan frame is obtained by redefining the metric using a positive field-dependent function $F(\f)$, with
\be \label{transf:metric}
g_{\mu \nu} = F(\f) g_{J  \mu \nu}.
\ee
The corresponding action is given by
\be \label{eq:Jaction}
S_J = \int \d^4 x \sqrt{-g_J} \left[F(\F) \frac{R_J}{2} - \h k g_J^{\mu \nu }\p_{\mu} \F \p_{\nu} \F - V_J(\F) + {\cal L}_m(\psi,g_{J\mu\nu})\right],
\ee
where we have included the possibility for the kinetic term to be of the ``wrong'' sign by keeping the prefactor $k$ unspecified for now. Note that we have added a matter Lagrangian to the model, where we assume that the matter couples to the Jordan frame metric, with the consequence that the Jordan frame metric may be regarded as the physical metric. The Jordan frame scalar field $\Phi$ is defined via
\be \label{transf:phi}
\frac{\d \F}{\d \f} = \sqrt{\frac{F}{k} \left(1- \frac{3}{2} \frac{F_{,\f}^{\,2}}{F^2}\right)}
\ee
and the potential becomes
\be \label{eq:V_J}
V_J (\F) = F(\f)^2 V(\f).
\ee
From the metric transformation (\ref{transf:metric}), we can immediately deduce the transformation of the scale factor,
\be \label{eq:aEJ}
a = \sqrt{F} a_J.
\ee
The transformation of the $00$-component of the metric is absorbed into the coordinate time interval,
\be
\d t = \sqrt{F} \d t_J,
\ee
such that the line element transforms as $\d s^2 = F(\phi) \d s_J^2.$ Moreover, by differentiating the scale factor w.r.t $\d t$, we can determine the Hubble parameter
\be \label{eq:H_E}
H \equiv \frac{a_{,t}}{a} = \frac{1}{\sqrt{F}} \left( H_J +\frac{F_{,t_J}}{2 F}\right),
\ee
where the Hubble parameter in Jordan frame is given by $H_J\equiv \frac{a_{J, t_J}}{a_J}$.

\subsection{A specific transformation}
We will now specialise to the following ansatz
\be \label{eq:F}
F(\f) = \xi \F^2 = e^{c \g \f},
\ee
which is inspired by the dilaton coupling in string theory, see for example \cite{Blumenhagen:2013fgp}, and has been used for instance in \cite{Li:2014qwa,Domenech:2015qoa}. This type of non-minimal coupling is also known as induced gravity \cite{Finelli:2007wb}; see e.g. \cite{Barrow:1990nv,Amendola:1990nn,Kamenshchik:2011yc} for related studies. %
Plugging in the solution for $\f$ from (\ref{eq:phi}), we can now integrate $\d t$
to find the relationship between the times in the two frames, yielding
\be \label{eq:transfts}
\frac{t_J}{t_{J,0}} = \left(\frac{t}{t_0}\right)^{1-\g},
\ee
where
\be
t_{J,0} = \frac{t_0}{1-\g}.
\ee
Using this result, we can calculate the scale factor in the Jordan frame from (\ref{eq:aEJ})
\be \label{eq:aJ}
a_J = a_0 \left(\frac{t}{t_0} \right)^{\frac{1-\ep \g}{\ep}} =a_0 \left(\frac{t_J}{t_{J,0}} \right)^{\frac{1-\ep \g}{\ep (1-\g)}}.
\ee
In order to obtain accelerated expansion, the $t_J$-exponent has to be larger than 1,
\be \label{constr:tjexp}
\frac{1-\ep \g}{\ep (1-\g)} > 1.
\ee
Moreover, an ekpyrotic phase in the Einstein frame has $\ep > 3$. From (\ref{constr:tjexp}), we see that for $\g < 1$ the denominator is positive and hence we would need $\ep < 1$, which cannot be satisfied for our case.
We conclude that to realise a phase of accelerated expansion in Jordan frame (from an ekpyrotic phase in Einstein frame), we need
\be \label{constraint:gamma}
\boxed{\g > 1}.
\ee

Another constraint is obtained from the relationship between the fields, given by the transformation in (\ref{transf:phi}) and the ansatz we have chosen for $F$ in (\ref{eq:F}).
Plugging in the latter into the first and integrating, we get
\be \label{eq:Phiphi}
\F = \frac{1}{\sqrt{\xi}} e^{^{c \g \f} /_2},
\ee
where the parameter $\xi$ is now determined in terms of $c = \sqrt{2\ep}$, $\g$ and $k$ and given as
\be \label{eq:xi}
\xi = \frac{c^2 \g^2 k}{4 - 6 c^2 \g^2}, 
\ee
or alternatively,
\be \label{eq:epsilon_xi}
\ep = \frac{2 \xi}{\g^2 \left( 6 \xi + k \right)}.
\ee
The parameter $\xi$ has to be positive for the gravity term in the Jordan frame action to be positive. A negative $\xi$ would lead to tensor ghosts.
Thus we need
\be
\xi >0  \;\;\;\;\;\;\; \iff \;\;\;\;\;\;\; k < 0 
\ee
since $\g > 1$ and $\epsilon > 3.$ Hence we see that we need the kinetic term for the scalar field to have the opposite of the usual sign, and we set
\be
\boxed{k= -1}.
\ee
Note that this ``wrong'' sign does not lead to ghosts, as there are additional contributions from the scalar-tensor coupling to the fluctuations of $\Phi,$ and these additional contributions render the total fluctuation positive (as we will show more explicitly in section \ref{sec:pertns}). With the above choice of $k$ we then also obtain a bound on the parameter $\xi$\footnote{In the language of Brans-Dicke scalar-tensor gravity, this condition translates to $\omega_{\text{BD}} > - 3/2$.},
\be \label{eq:xibound}
\boxed{\xi > \frac{1}{6}}.
\ee
The Jordan frame potential can be reexpressed in terms of $\F$ as
\be \label{eq:V_J}
V_J(\F) = F^2(\f) V(\f) = - V_0 e^{\left( 2 \g - 1 \right)c\f} = - V_{J,0}  \F^{4 - 2/\g},
\ee
where we have defined $V_{J,0} \equiv V_0 \xi^{2 - 1/\g}$. The negative exponential of the ekpyrotic phase gets transformed into a \emph{negative} power-law potential. We thus see that it is possible to obtain a phase of accelerated expansion in the presence of a negative potential in Jordan frame, starting from ekpyrosis in Einstein frame together with the conditions $\g > 1$, $k=-1$, and $\xi > 1/6$. We will refer to this new phase of accelerated expansion as the conflationary phase.

\subsection{Equations of motion in Jordan frame}

Varying the action (\ref{eq:Jaction}) w.r.t. the Jordan frame metric and scalar field, we obtain the Friedmann equations and the equation of motion for the scalar field $\F$:
\bea \label{eq:JFriedmann1}
&& 3H_J^2 F + 3 H_J F_{,t_J} = \h k \F_{,t_J}^2 + V_J, \\
\label{eq:JFriedmann2}
&& 2 F H_{J,t_J} + k \F_{,t_J}^2 - H_J F_{,t_J} + F_{,t_J t_J} =0, \\
\label{eq:JPhieom}
&& \F_{,t_J t_J} + 3 H_J \F_{,t_J} - \frac{3 F_{,\F}}{k} \left( H_{J,t_J} + 2 H_J^2 \right) + \frac{V_{J,\F}}{k} =0.
\eea
The first Friedmann equation (\ref{eq:JFriedmann1}) can be solved for the Hubble parameter,
\be \label{eq:HJ12}
H_{J} = - \frac{F_{,t_J}}{2 F} \pm \sqrt{\frac{F_{,t_J}^2}{4 F^2} + \frac{k}{6 F} \F_{,t_J}^2 + \frac{1}{3 F} V_J}.
\ee
$H_J$ will give two positive solutions as the square root is always less than $- \frac{F_{,t_J}}{2 F}>0$, since $k, V_J < 0$.
To determine the solution that corresponds to contraction in Einstein frame, we note that the Hubble parameter in Einstein frame given in (\ref{eq:H_E}) has to be negative.
Hence, we have to pick out the solution for $H_J$ which satisfies
\be \label{eq:transfHs}
H_J <  - \frac{F_{,t_J}}{2 F}.
\ee
This is exactly the term to which the square root is added or subtracted in (\ref{eq:HJ12}), and thus we have to choose the latter:
\be \label{eq:HJ}
H_{J} = - \frac{F_{,t_J}}{2 F} - \sqrt{\frac{F_{,t_J}^2}{4 F^2} + \frac{k}{6 F} \F_{,t_J}^2 + \frac{1}{3 F} V_J}.
\ee

We can rewrite $\F$ as a function of Jordan frame time, $t_J$, using equations (\ref{eq:phi}) and (\ref{eq:transfts}),
\be \label{eq:PhitJ}
\F (t_J) = \frac{1}{\sqrt{\xi}} \left(\frac{t_J}{t_{J,0}}\right)^{\frac{\g}{1-\g}}.
\ee
We can then determine the quantity $V_J/\F_{,t_J}^2$ using Eq. (\ref{eq:V_J}), obtaining
\be \label{eq:VJPhitJ}
\frac{V_J}{\F_{,t_J}^2} = \frac{\ep - 3}{\ep \left( 2 - 6 \ep \g^2 \right)}\,.
\ee
This combination is (non-trivially) time-independent, and hence once it is satisfied for the initial conditions of a particular solution it will hold at any time. This equation will be useful in setting the initial conditions for specific numerical examples, as will be done in the next section.

\subsection{Initial conditions and evolution with a shifted potential}
\begin{figure}[htbp]
\begin{minipage}{0.5\textwidth}
\includegraphics[width=0.92\textwidth]{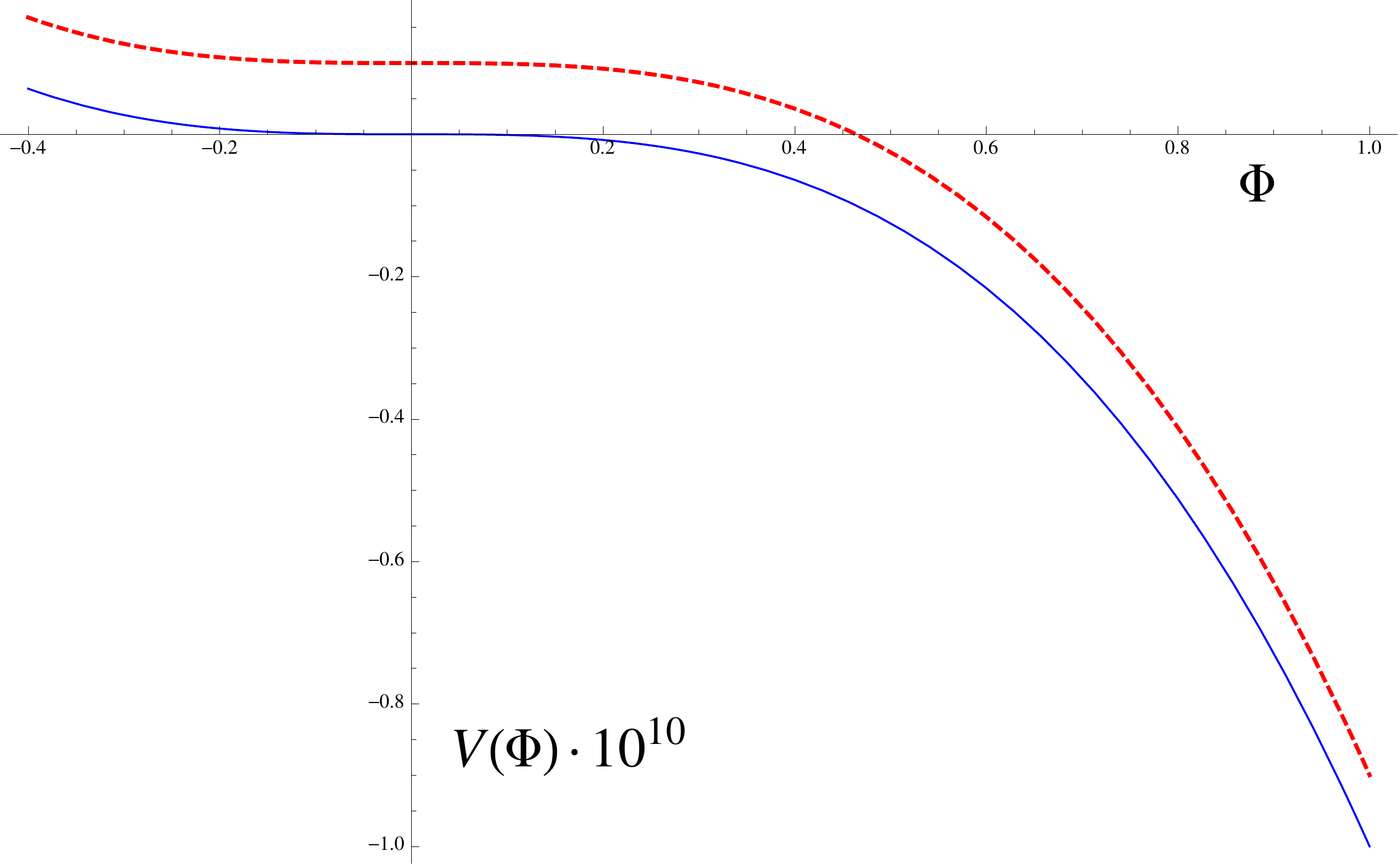}
\end{minipage}%
\begin{minipage}{0.5\textwidth}
\includegraphics[width=0.92\textwidth]{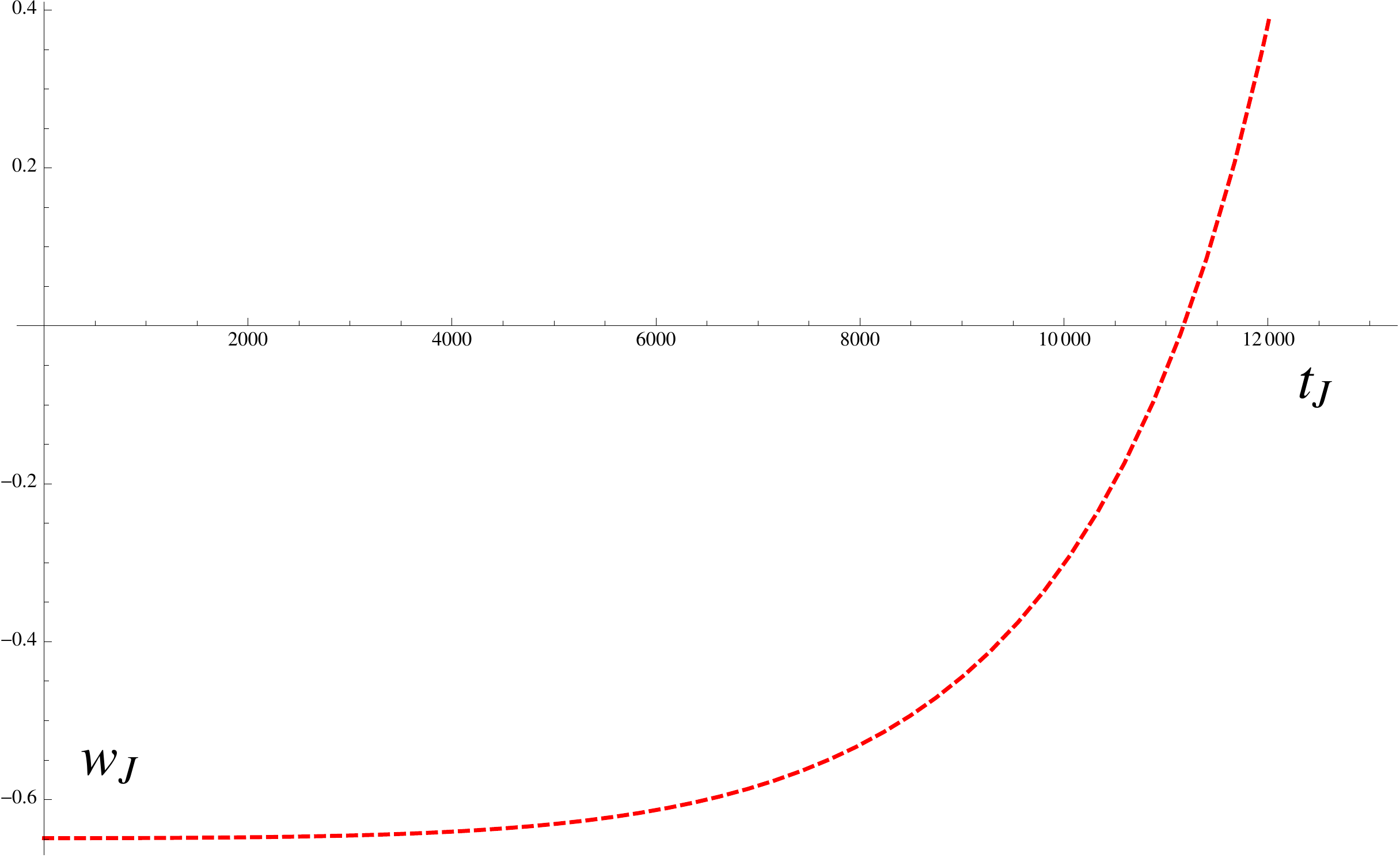}
\end{minipage}%
\caption{ \label{fig:J_potl} {\it Left:} The original Jordan frame potential $V_J$ is shown in blue, the shifted potential $U_J$ in dashed red. {\it Right:} The equation of state in Jordan frame, for the shifted potential.} 
\end{figure}

\begin{figure}[htbp]
	\begin{minipage}{0.65\textwidth}
		\includegraphics[width=1\textwidth]{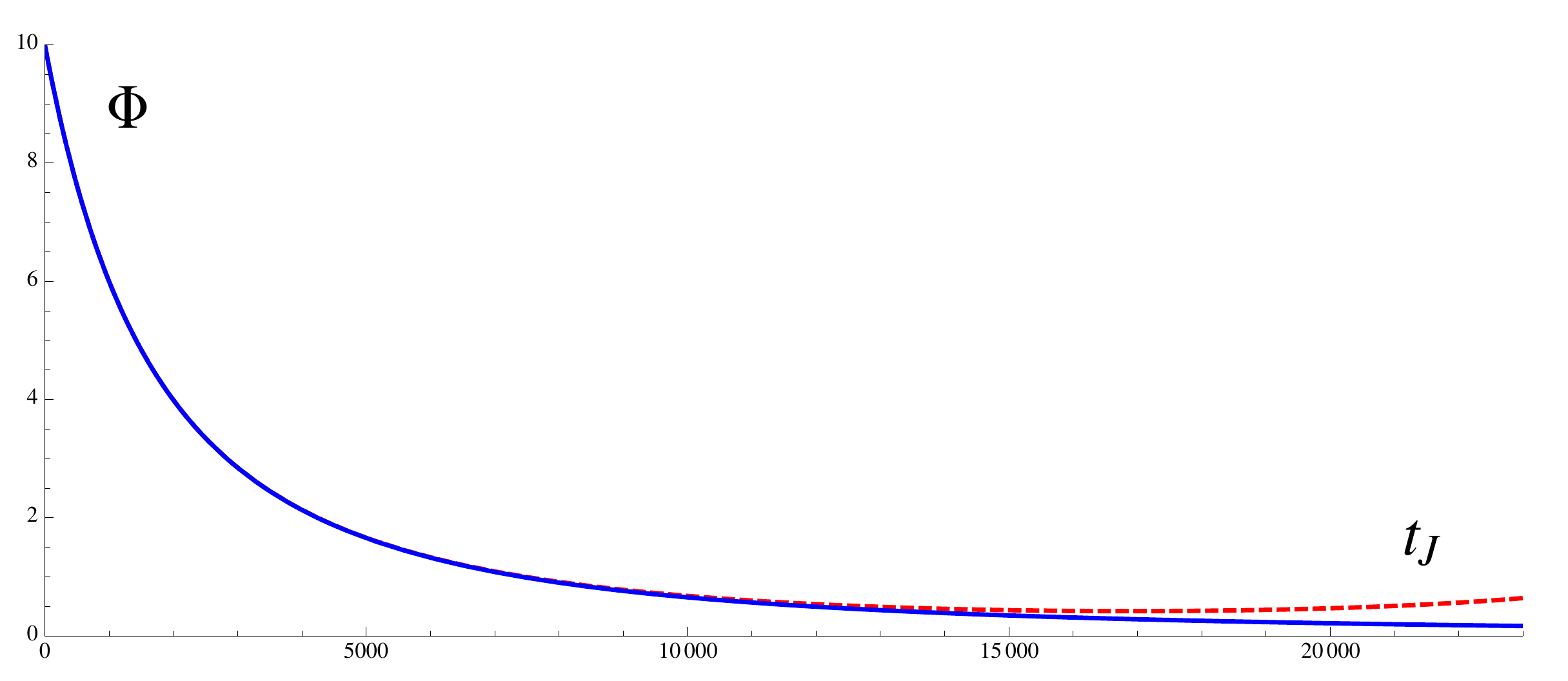}
		\includegraphics[width=1\textwidth]{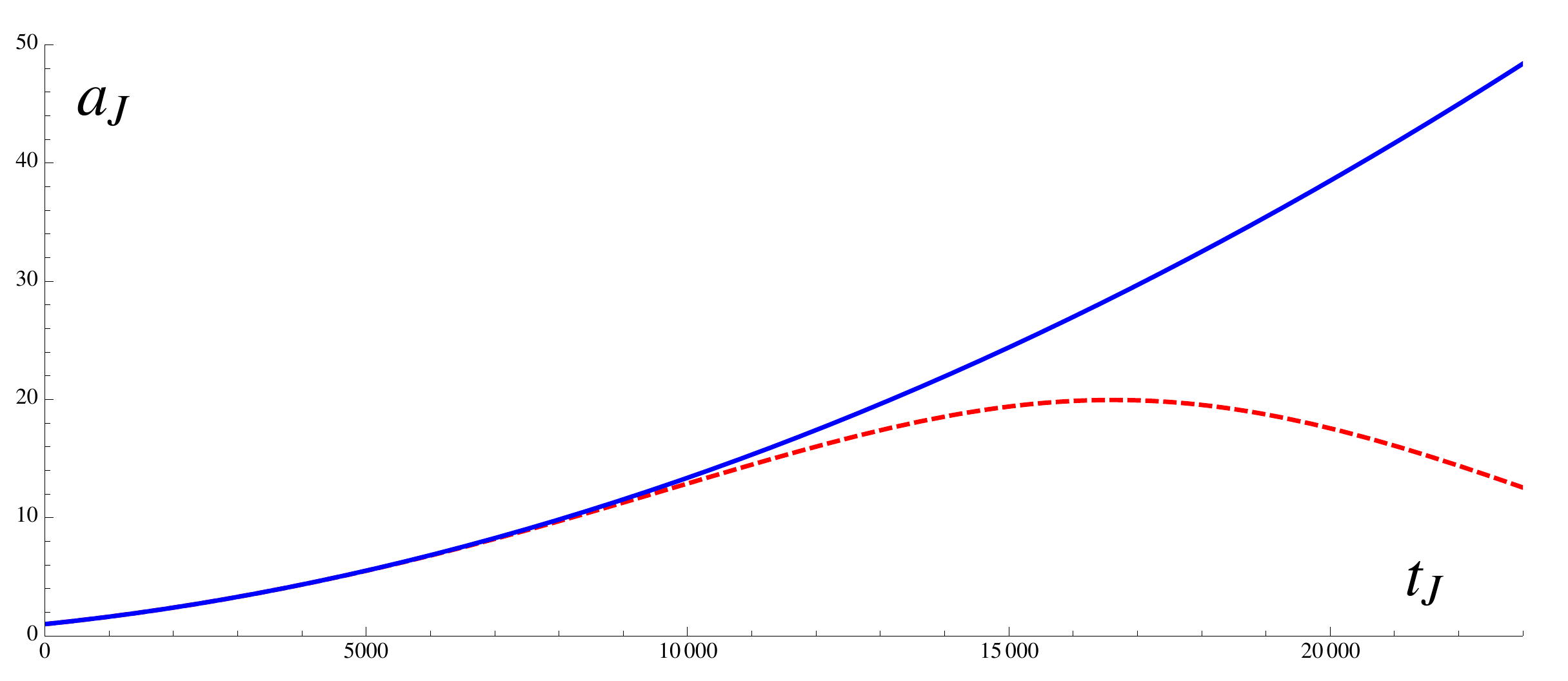}
	\end{minipage}%
	\caption{Scalar field and scale factor in Jordan frame: the blue curves show the transformed ekpyrotic scaling solution and the red dashed curves correspond to the field evolutions in the shifted potential. }
	\protect
	\label{fig:J_Phi_a}
\end{figure}

In this subsection we verify that our construction indeed leads to accelerated expansion in Jordan frame. 
We choose the parameters $\epsilon=10$ and $\g=2$ leading to a negative $\F^3$ potential in Jordan frame -- see Fig. \ref{fig:J_potl}. For an initial field value of $\F(t_{beg})=10$ and $V_{J,0}= 10^{-10}$, we require an initial field velocity (using equations (\ref{eq:V_J}) and (\ref{eq:VJPhitJ})) of
$| \F_{,t_J} | \approx 5.83 \cdot 10^{-3}$. Furthermore we set $a_J(t_{beg})=1$. Numerical solutions for the scale factor and scalar field are shown in Fig. \ref{fig:J_Phi_a}, where the blue curves indeed reproduce the conflationary transform of the ekpyrotic scaling solution.

Note that it follows from equations (\ref{eq:aJ}) and (\ref{eq:PhitJ}) -- similarly to inflationary models -- that there is a spacetime singularity at $t_J = 0, \, a_J =0, \, \F = \infty$, which should be resolved in a more complete theory. Either the effective description might break down at that time, or we never reach such times in a more complete (cyclic) embedding of the theory. We leave such considerations for future work.

Eventually, the conflationary phase has to come to an end. As a first attempt at a graceful exit we shift the potential in Jordan frame by a small amount $V_1$ (it will turn out that this simple modification is too naive and we will improve on it in the next subsection),%
\be
U_J(\F) = V_J(\F) + V_1\,.
\ee
The shifted potential, with $V_1=\frac{V_{J,0}}{10},$ is plotted in Fig. \ref{fig:J_potl}. The corresponding evolution of the scalar field $\F$ and the scale factor in Jordan frame are now shown as the red dashed curves in Fig. \ref{fig:J_Phi_a}, while the equation of state is plotted in the right panel of Fig. $\!$\ref{fig:J_potl}. The conflationary phase lasts until $t_J \approx 10000$ when the equation of state grows larger than $w_J=-1/3,$ and accelerated expansion ends. The scalar field continues on to about $\F \approx 0.4$ and then rolls back down the potential. Meanwhile, the scale factor reaches a maximum value and starts re-contracting. This re-contraction in Jordan frame is unavoidable: from equation (\ref{eq:HJ}), bearing in mind that $F_{,t_J}<0,$ it becomes clear that whenever $\rho_J=\frac{k}{2} \F_{,t_J}^2 + V_J=0$ we have $H_J =0$ resulting in a re-contraction in Jordan frame. Given that we start out with both a negative kinetic term ($k=-1$) and a negative potential, but then want to reach positive potential values, means that we will necessarily pass through $\rho_J=0$ as the scalar field slows down. It is clear that a shift in the Jordan frame potential is not sufficient for a graceful exit -- more elaborate dynamics are needed to avoid collapse. %
One might imagine that the scalar field could stabilise at a positive value of the potential. It could then either stay there and act as dark energy, or decay such that reheating would take place. Once the scalar field stabilises, the Einstein and Jordan frame descriptions become essentially equivalent\footnote{When the scalar field is constant, the two frames are equivalent. However, when the scalar field is perturbed, then fluctuations in the Jordan frame will still feel the direct coupling to gravity.}. However, this means that the scale factor will only revert to expansion if a bounce also occurs in Einstein frame. This motivates us to extend the present model by including dynamics that can cause a smooth bounce to occur after the ekpyrotic phase.

\begin{figure}
	\centering
	\includegraphics[width=0.5\textwidth]{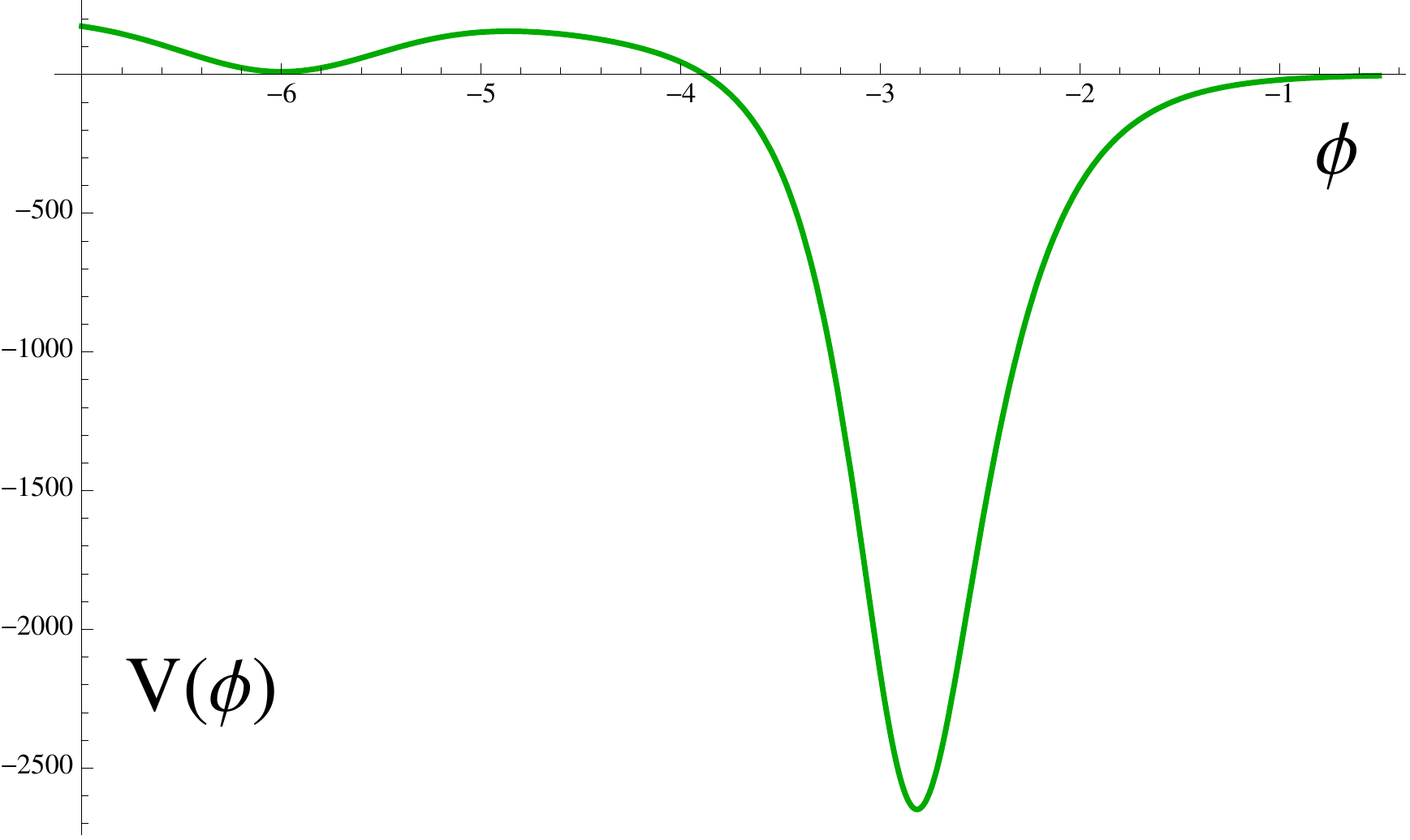}
	\caption{The Einstein frame scalar potential used in the bounce model \eqref{eq:actionp}.}
	\protect
	\label{fig:e_bounce_pot}
\end{figure}

\subsection{Transforming an Einstein frame bounce} \label{sec:E_bounce}

\begin{figure}[htbp]
	\begin{minipage}{0.5\textwidth}
		\includegraphics[width=1.0\textwidth]{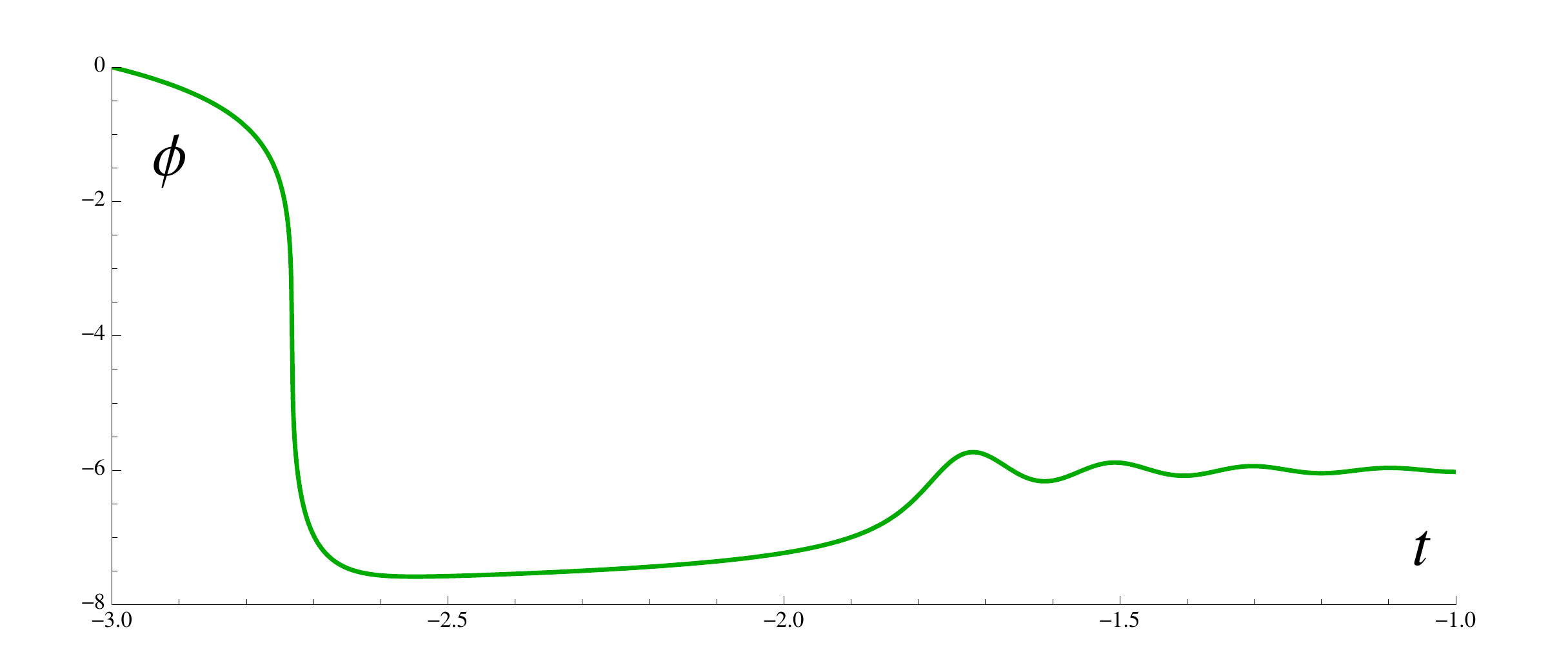}
		\includegraphics[width=1.0\textwidth]{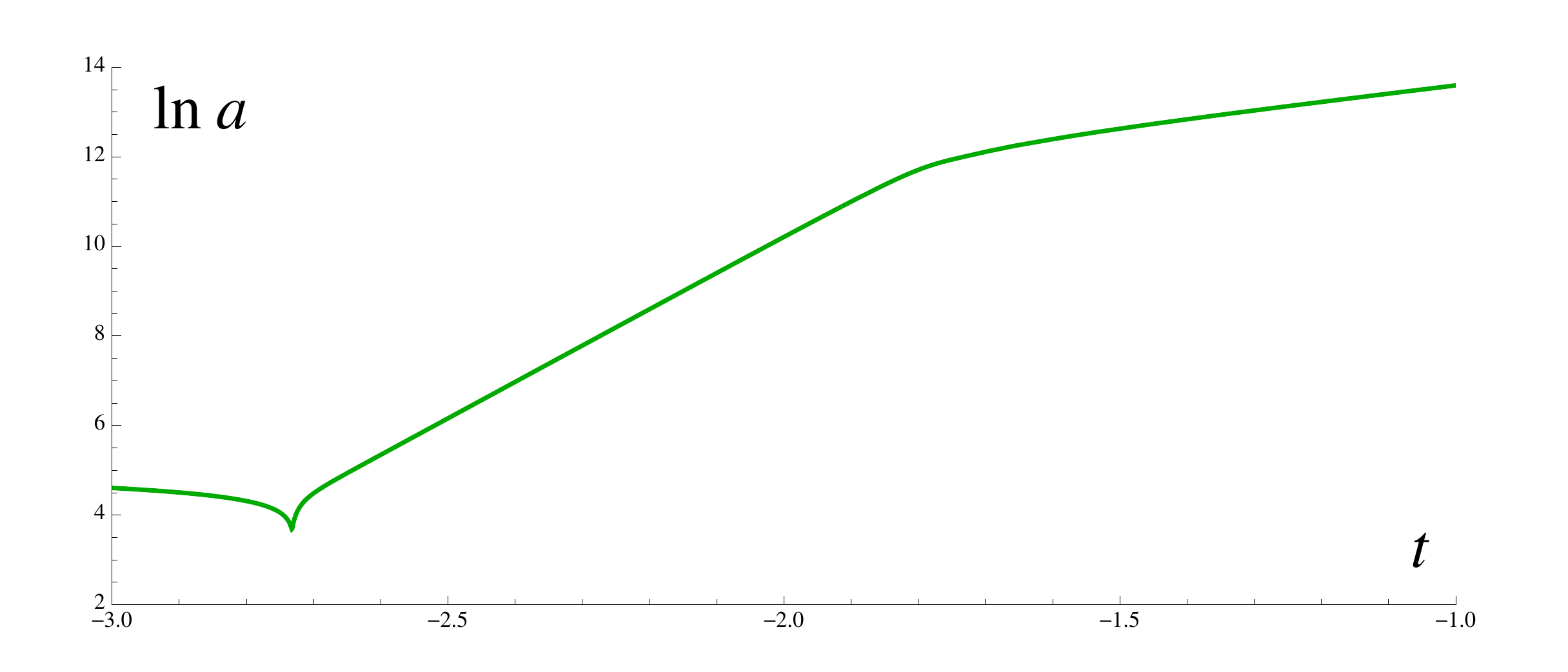}
	\end{minipage}%
	\begin{minipage}{0.5\textwidth}
		\includegraphics[width=1\textwidth , height=6.8cm]{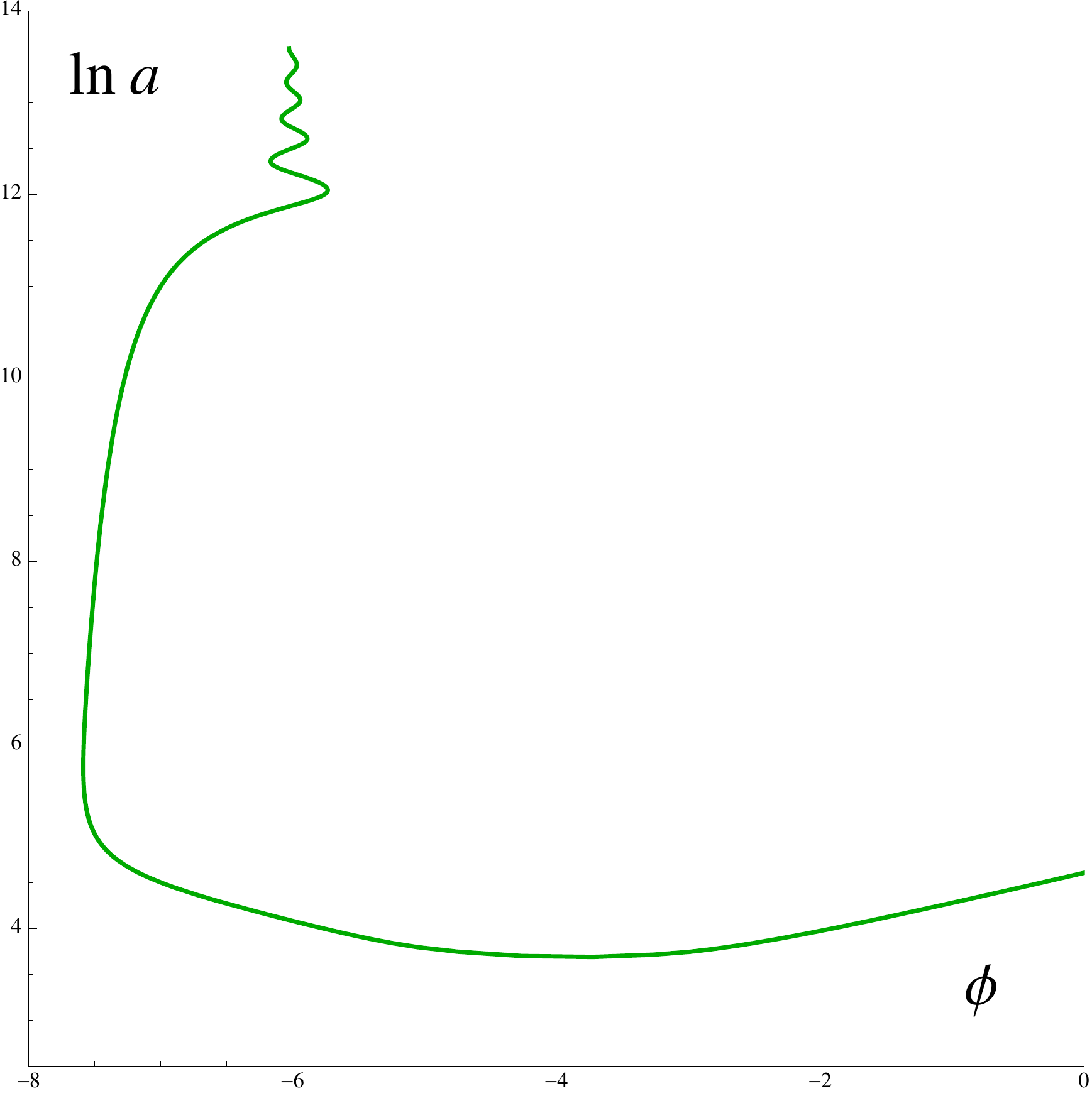}
	\end{minipage}%
	\caption{ \label{fig:num_E} {\it Left:} Scalar field and scale factor for the bounce solution in Einstein frame. {\it Right:} Parametric plot of the scalar field and scale factor in Einstein frame. This plot nicely illustrates the smoothness of the bounce.} 
\end{figure}

In ekpyrotic models, after the ekpyrotic contracting phase has come to an end the universe must bounce into an expanding hot big bang phase. Many ideas for bounces have been put forward, see e.g. \cite{Buchbinder:2007ad, Creminelli:2007aq, Qiu:2011cy, Easson:2011zy, Cai:2012va, Qiu:2015nha, Turok:2004gb, Bars:2011aa} -- here we will focus on a non-singular bounce achieved via a ghost condensate \cite{Koehn:2013upa, Lehners:2015efa}. This model has the advantage of being technically fairly simple, and, importantly, it is part of a class of models for which it has been demonstrated that long-wavelength perturbations are conserved through the bounce \cite{Xue:2013bva,Battarra:2014tga}. Moreover, it was shown in \cite{Koehn:2015vvy} (where the scale at which quantum corrections occur was calculated) that such models constitute healthy effective field theories. 
The action we will consider takes the form
\be \label{eq:actionp}
S = \int \d^4 x \sqrt{-g} \left[\frac{R}{2} + P(X,\f) \right]
\ee
with 
\be
P(X, \f)= K(\f)X+Q(\f)X^2 - V(\f)
\ee
and where $X\equiv -\frac{1}{2} g^{\mu \nu }\p_{\mu} \f \p_{\nu} \f$ denotes the ordinary kinetic term. The shape of the functions $K(\f)$ and $Q(\f)$ can be chosen in various ways. The important feature is that at a certain time (here at $\f=-4$) the higher derivative term is briefly turned on while the sign of the kinetic term changes. Moreover, we add a local minimum to the potential, as shown in Fig. \ref{fig:e_bounce_pot}: after the bounce the scalar field rolls into a dip in the potential where the scalar field stabilises and where reheating can occur. For specificity we will use the functions \cite{Lehners:2015efa}
\bea
K(\f) &=& 1-\frac{2}{\left (1+\frac{1}{2}(\f+4)^2\right)^2} \,, \\
Q(\f)&=&\frac{V_{0}}{\left (1+\frac{1}{2}(\f+4)^2\right)^2}\,, \\
V(\f) &=&-\frac{1}{e^{3\f}+e^{-4(\f+5)}}+100\left [ \left( 1 - \tanh(\f + 4)\right) \left(1 - 0.95 e^{-2(\f + 6)^2} \right)\right ]\,,
\eea
where compared to \cite{Lehners:2015efa} the theory has been rescaled according to $g_{\mu \nu} \to V_0^{1/2}g_{\mu \nu}$  which implies $K \to K$, $Q \to V_{0} Q$ and $V \to V_{0}^{-1} V$. The equations of motions obtained by varying the action (\ref{eq:actionp}) read
\bea
\nabla_\m \left( P_{,X} \nabla^{\m} \f\right)-P_{,\f}=0\\
3H^2=\rho \\
\dot H=-\frac{1}{2}(\rho+p) 
\eea
where the pressure and energy density are given by $p=P$ and $\rho=2XP_{,X}-P$. Note that $\dot{H}=-XP_{,X},$ which shows that the Hubble rate can increase (as is necessary for a bounce) when the ordinary kinetic term switches sign. The purpose of the $X^2$ term in the action is twofold: it allows the coefficient of the ordinary kinetic term to pass through zero, and it contributes to the fluctuations around the bounce solution in such a way as to avoid ghosts. The Einstein frame bounce solution is shown in Fig. \ref{fig:num_E}, where we have chosen the initial conditions $\f_0= 0$, $\dot\f_0 =-2.4555$, $a_{0}=100$ and have set $V_0=10^{-6}$ and $c=3$. The scalar field first rolls down the potential during the ekpyrotic phase. A bounce then occurs near $\f=-4$ due to the sign change of the kinetic term. After this, the universe starts expanding, the potential becomes positive and the scalar field rolls into the dip where it oscillates with decaying amplitude -- see Fig. \ref{fig:num_E}.

In the following we want to transform this bouncing solution into Jordan frame, in order to see how such a bounce translates into a graceful exit for the conflationary phase. The Ricci scalar transforms under our conformal transformation (\ref{transf:metric}) as \cite{Wald:1984r}
\be
R=\frac{1}{F} \left( R_J-6\Box_J \ln \sqrt{F} -6g_J^{\m \n}\p_\m \left (\ln {\sqrt F}\right) \p_\n \left (\ln {\sqrt F} \right) \right)\,,
\ee
where the second term contributes as a total derivative in the action. 
Note that the kinetic term transforms as
\be
X \equiv -\frac{1}{2} g^{\mu \nu }\p_{\mu} \f \p_{\nu} \f=-\frac{1}{2F} g_J^{\mu \nu }\p_{\mu} \f \p_{\nu} \f=-\frac{1}{2F} \left(\frac{\p \f}{\p \F} \right)^2 g_J^{\mu \nu }\p_{\mu} \F \p_{\nu }\F \equiv \frac{1}{F}\left(\frac{\p \f}{\p \F} \right)^2X_J.
\ee
Plugging everything into equation (\ref{eq:actionp}) yields the action in Jordan frame
\be
S_J = \int \d^4 x \sqrt{-g_J} \left [F(\Phi)\frac{R_J}{2} +P_J(X_J,\Phi) \right]\,,
\ee
where we have defined the new functions in Jordan frame as
\bea
P_J&\equiv&K_J X_J+Q_JX^2_J-V_J,\\
K_J&\equiv&F \left(K-\frac{3}{2}\frac{F^{\,2}_{,\f}}{F^2} \right)\left(\frac{\p \f}{\p \F} \right)^2 = 4 \xi\left(\frac{K}{c^2 \g^2} -\frac{3}{2} \right), \\
Q_J&\equiv&Q \left(\frac{\p \f}{\p \F} \right)^4 = \frac{16}{c^4 \g^4 \F^4}Q, \\
V_J&\equiv&F^2V = \xi^2 \F^2V,
\eea
where we have used 
\be
\frac{\p \f}{\p \F}=\frac{2}{c \g \F}  \quad \text{and} \quad F(\F)=\xi \F^2.
\ee
Thus the equations of motions in Jordan frame are given by
\bea
\nabla_\m \left( P_{J,X} \nabla^{\m} \F\right)=P_{J,\F}+\frac{1}{2}R_J F_{,\F}\\
3FH_J^2+3H_J F_{,t_J}=\rho_J \\
\rho_J+p_J +2F H_{J,t_J}-H_J F_{,t_J}+F_{,t_J t_J}=0
\eea
with the effective energy density $\rho_J=2X_JP_{J,X}-P_J$ and effective pressure $p_J=P_J$.

\begin{figure}[htbp]
	\includegraphics[width=0.6\textwidth]{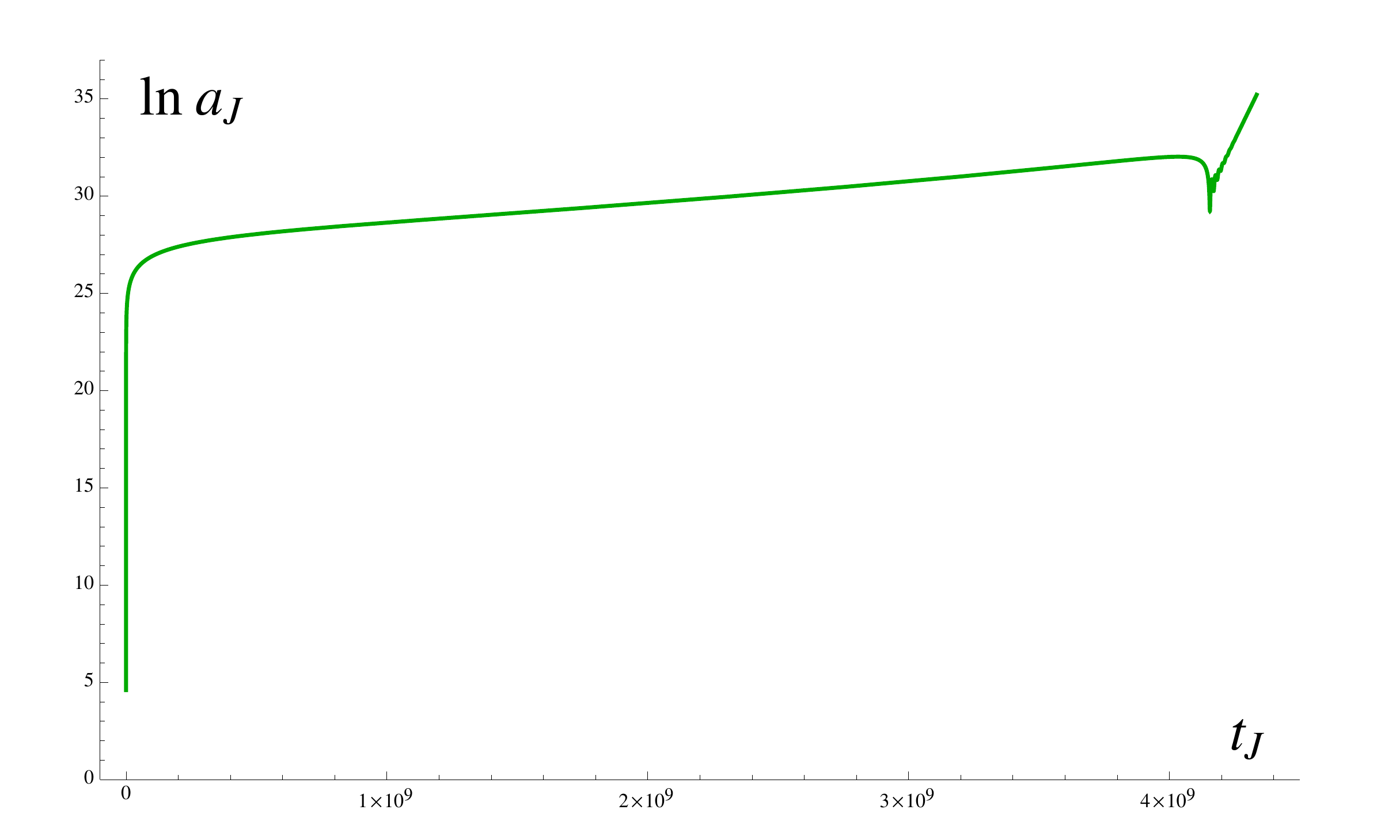}
	\caption{ \label{fig:num_J_full} Full evolution of the scale factor for the transformed solution in Jordan frame. The conflationary phase lasts while the scalar field rolls up the potential towards $\F \sim 10^{-9}$. During this period the scale factor increases by many orders of magnitude. During the exit of the conflationary phase the scale factor and scalar field undergo non-trivial evolution which is hard to see in the present figure and is shown in detail in Fig. \ref{fig:num_J}} 
\end{figure}%

\begin{figure}[htbp]
	\begin{minipage}{0.5\textwidth}
		\includegraphics[width=1\textwidth]{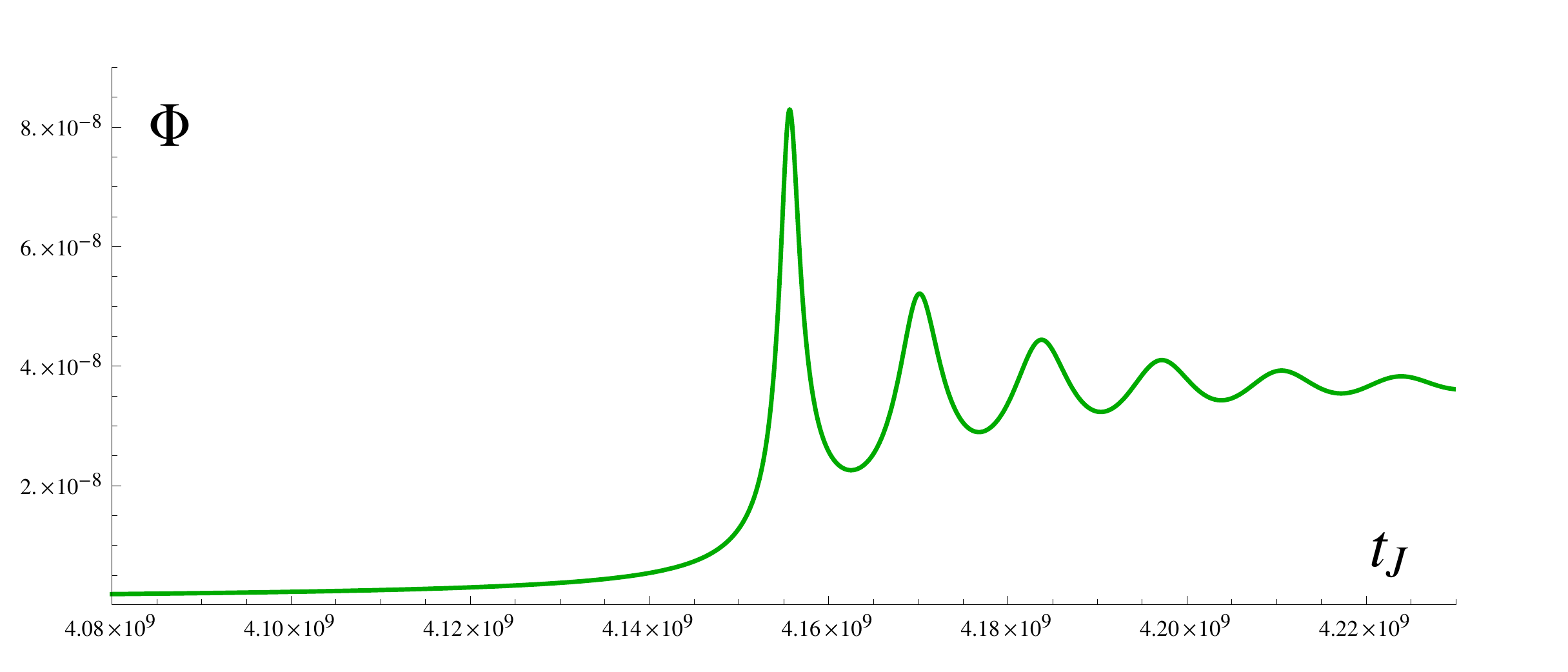}
		\includegraphics[width=1\textwidth]{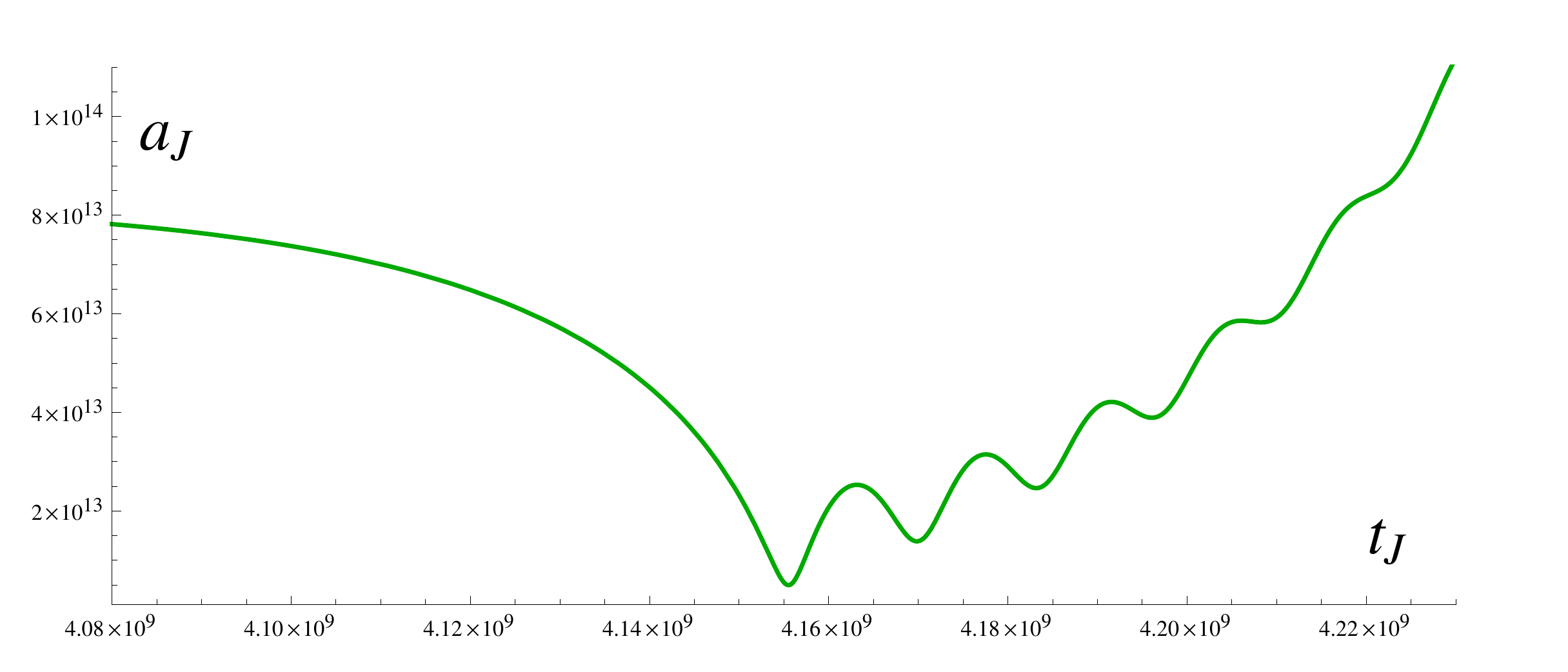}
	\end{minipage}%
	\begin{minipage}{0.5\textwidth}
		\includegraphics[width=1\textwidth,height=6.7cm]{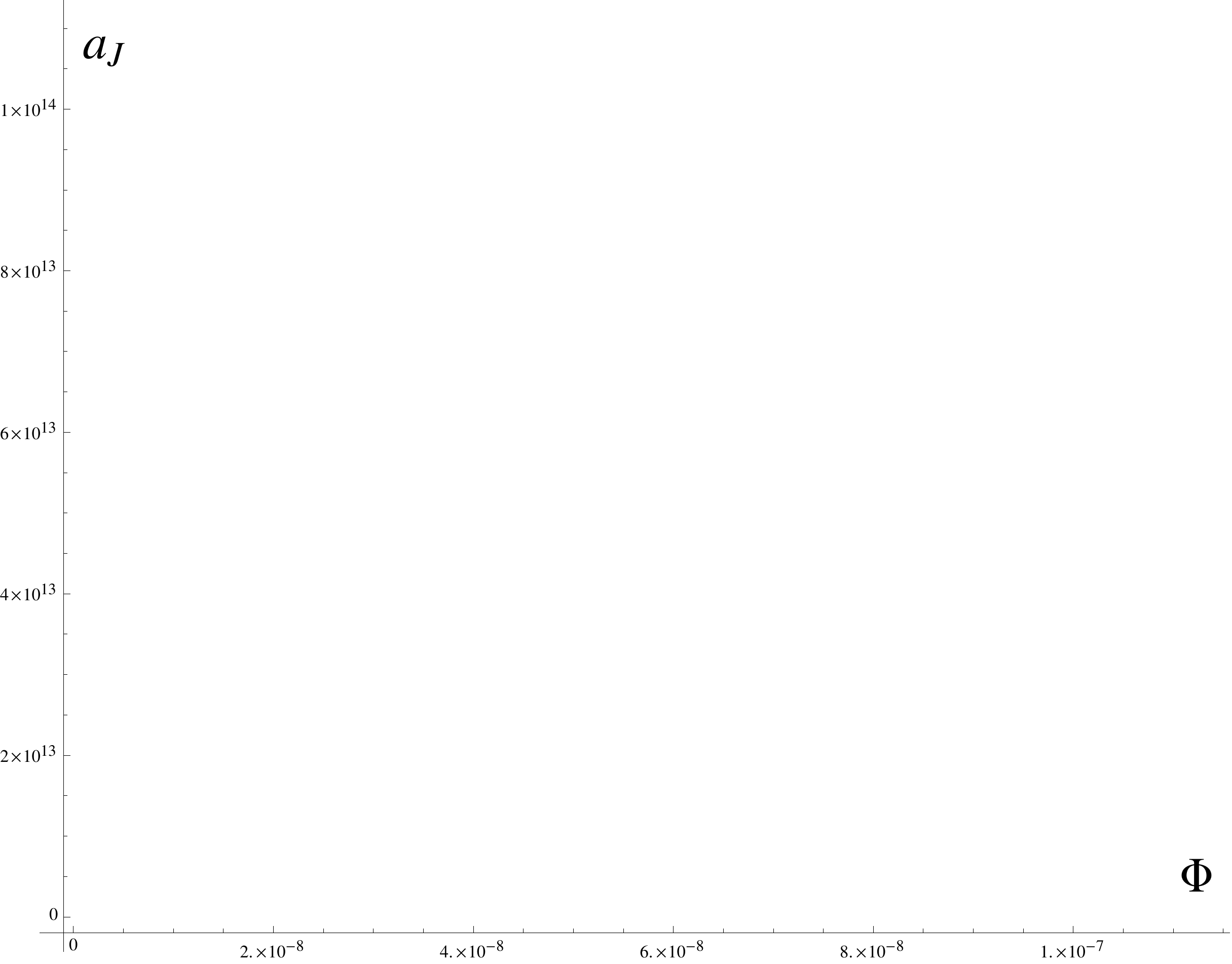}
	\end{minipage}%
	\caption{ \label{fig:num_J} {\it Left:} Scalar field and scale factor for the transformed solution in Jordan frame towards the end of the evolution. {\it Right:} Parametric plot of the scalar field and scale factor in Jordan frame. Note that initially the scalar field decreases its value very rapidly. Later on as the scalar field stabilises, the scale factor goes through oscillations, but eventually increases monotonically.} 
\end{figure}

The conflationary solution is shown in Fig. \ref{fig:num_J_full}. The scalar field $\F$ rolls up the approximately $-\Phi^3$ potential with decreasing velocity. It starts out at $\F_0 = 2.4267$ and very quickly decreases to a field value $\F \sim 10^{-9}$ where it stays for a long time. By this time, the bounce in Einstein frame has already taken place, but interestingly it leads to nothing dramatic in Jordan frame -- the universe simply keeps expanding and the scalar field keeps decreasing. The more interesting dynamics in Jordan frame occurs later, see Fig.~\ref{fig:num_J}. As we have already discussed, the universe re-contracts for $\rho_J=0$. The potential energy increases to positive values (in this model accelerated expansion ends as the potential becomes positive!) and the kinetic term decreases leading to a re-contraction at $t_J \approx 4.05 \cdot 10^{9}$. The re-contraction $H_J<0$ leads to an increased scalar field velocity, allowing the scalar field to roll over the potential barrier and into the dip, where it starts oscillating around the minimum, eventually settling at the bottom. The Hubble rate $H_J$ changes sign each time the energy density passes through zero, so that the scale factor oscillates together with the scalar field. Once the scalar field is settled, continuous expansion occurs. Note that these oscillations of the scale factor do not correspond to a violation of the null energy condition -- they are simply due to the coupling between the scalar field and gravity in Jordan frame. It would be interesting to see whether reheating might speed up the settling down of the scalar field -- we leave such an analysis for future work.

\section{Perturbations}

It is known that under a conformal transformation of the metric perturbations are unaffected. Thus we know what kind of cosmological perturbations our model leads to: during the ekpyrotic phase, both adiabatic scalar fluctuations and tensor perturbations obtain a blue spectrum and are not amplified. However, with the inclusion of a second scalar field, nearly scale-invariant entropy perturbations can be generated first, which can then be converted into adiabatic scalar curvature fluctuations at the end of the ekpyrotic phase. Translated into the conflationary framework of the Jordan frame, these results are nevertheless surprising: they imply that we have a phase of accelerated expansion during which adiabatic perturbations as well as tensor fluctuations have a spectrum very far from scale-invariance, and moreover they are not amplified. It is thus instructive to calculate these perturbations explicitly in this frame, which is what we will do next. In the following subsection, we will also describe the entropic mechanism from the point of view of the Jordan frame. Throughout this section, we will use the notation that a prime denotes a derivative w.r.t. conformal time $\tau,$ which is equal in both frames as $\d t/a = \d t_J/a_J$.

\subsection{Perturbations for a single scalar field}\label{sec:pertns}

As has been calculated for instance in \cite{Li:2014qwa}, the quadratic action for the comoving curvature perturbation $\zeta_J$ in Jordan frame is given by 
\be
S_J^{(2)} = \h \int \d^4 x  \frac{a_J^2 \F'^2}{\left(\cH_J+\frac{\F'}{\F}\right)^2}  \left(  6 \xi -1 \right) \left(\z_J'^2 -\left(\p_i \z_J \right)^2 \right),
\ee
where we have assumed $F(\Phi) = \xi \Phi^2.$ The absence of ghost fluctuations can thus be seen to translate into the requirement
\be
\xi > \frac{1}{6},
\ee
which is the same condition on $\xi$ that we had discovered before in Eq. (\ref{eq:xibound}). We can define
\be
z_J^2=\frac{a_J^2 \F'^2}{\left(\cH_J+\frac{\F'}{\F}\right)^2}  \left(  6 \xi -1 \right)\,,
\ee
so that for the canonically normalised Mukhanov-Sasaki variable  $v_{J}=z_J \zeta_J$ we obtain the mode equation in standard form, namely
\be
v_{Jk}''+ \left(k^2-\frac{z_{J}''}{z_{J}} \right) v_{Jk}=0\,.
\ee 
Note however that $z_J$ does \emph{not} have the usual form $\sim a_J\Phi^{\prime}/\cH_J,$ but the denominator contains an extra contribution from the scalar field. This contribution  is crucial, as it implies that the usual intuition gained from studying inflationary models in Einstein frame is not applicable here. For the conflationary transform of the ekpyrotic scaling solution we have
\be \label{eq:aJ2}
a_J (t_J)=a_0 \left(\frac{t_J}{t_{J,0}} \right)^{\frac{1-\ep \g}{\ep (1-\g)}}, \quad \F (t_J) = \frac{1}{\sqrt{\xi}} \left(\frac{t_J}{t_{J,0}}\right)^{\frac{\g}{1-\g}},
\ee
while the relationship between physical time and conformal time is given by
\be \label{eq:conft}
t_J \sim (-\tau)^{\frac{\ep(1-\g)}{\ep-1}}\,.
\ee
These relations imply that $z_J(\tau) \sim (-\tau)^{1/(\ep-1)}$ which leads to
\be
\frac{z_J''}{z_J}=\frac{2-\ep}{(\ep-1)^2}\frac{1}{\tau^2}\,.
\ee
Imposing Bunch-Davies boundary conditions in the far past selects the solution (given here up to a phase)
\be
v_{J k} = \sqrt{-\frac{\pi}{4}\tau} H_{\nu}^{(1)}(-k\tau)\,,
\ee
where $H_\nu^{(1)}$ is a Hankel function of the first kind with index $\nu = \h - \frac{1}{\ep -1}$. This leads to a scalar spectral index
\be
n_{\zeta}-1 \equiv 3-2\nu=3-\left|\frac{\ep-3}{\ep-1}\right|,
\ee
where $\epsilon$ corresponds to the Einstein frame slow-roll/fast-roll parameter. Here $\epsilon > 3$ and thus the (blue) spectrum is always between $3 < n_\zeta < 4,$ i.e. the spectrum is identical to that of the adiabatic perturbation during an ekpyrotic phase, as expected \cite{Lyth:2001pf}. 

The calculation of the (transverse, traceless) tensor perturbations $\gamma_{J ij}$ proceeds in an analogous fashion. Their quadratic action is given by
\be
S_J=- \frac{1}{8}\int d^4x F(\F)\sqrt{g_J}g_J^{\mu \nu}\p_{\mu}\gamma_{Jij}\p_{\nu}\gamma_{Jij}\,.
\ee
Writing the canonically normalised perturbations as $h \; \! \epsilon_{ij}\equiv z_T \gamma_{J ij},$ where $\epsilon_{ij}$ is a polarisation tensor and where $z_T^2= F(\Phi) a_J^2,$ the mode equation in Fourier space again takes on the usual form
\be
h_{k}''+\left(k^2-\frac{z_T''}{z_T}\right) h_{k}=0\,,
\ee
except that here $z_T$ is not just given by the scale factor but involves the scalar field too. In fact $z_T \propto \Phi a_J \propto (-\tau)^{1/(\epsilon -1)}$ and thus $z_T \propto z_J.$ The spectral index comes out as
\be
n_T \equiv 3- \left|\frac{\ep-3}{\ep-1}\right|\,,
\ee
which is the same blue spectrum as that obtained during an ekpyrotic phase, as it must. 

These simple calculations have an important consequence. In the limit where $|k\tau|\ll1,$ which corresponds to the late-time/large-scale limit, the adiabatic scalar and tensor mode functions and momenta have the asymptotic behaviour \cite{Tseng:2012qd,Battarra:2013cha} $(\nu = \frac{1}{2} - \frac{1}{\epsilon - 1})$
\bea \label{eq:scalartensormodes}
v_{J k}\,, \, h_{k}\, &\approx& \frac{\pi^{\frac{1}{2}}k^\nu}{2^{\nu+1}\Gamma(\nu + 1)}(-\tau)^{1 - \frac{1}{\epsilon -1 }}+ i\left[ -\frac{2^{\nu-1}\Gamma(\nu)}{\pi^{\frac{1}{2}}k^\nu} (-\tau)^{\frac{1}{\epsilon -1 }} - \frac{\cos(\pi\nu)k^\nu \Gamma(-\nu)}{\pi^{\frac{1}{2}}2^{\nu+1}} (-\tau)^{1- \frac{1}{\epsilon -1 }}\right]\quad \\
\pi_{v,h} &\approx& \frac{\nu \pi^{\frac{1}{2}}k^\nu}{2^{\nu}} \left(- \frac{1}{\Gamma(\nu + 1)} + i \frac{\cos(\pi\nu)\Gamma(-\nu)}{\pi} \right) (-\tau)^{- \frac{1}{\epsilon -1 }} \,, 
\eea
where the momenta are defined as $\pi_v = v^\prime_J - \frac{z_J^\prime}{z_J}v_J,\, \pi_h = h^\prime - \frac{z_T^\prime}{z_T}h.$ When $\epsilon$ is large, $\epsilon \gg 1$ and consequently $\nu \approx \frac{1}{2},$ these expressions tend to the Minkowski space mode functions and momenta. In this limit, hardly any amplification nor squeezing of the perturbations occurs. This is in stark contrast with standard models of inflation where $\ep < 1$ and where the second term on the right hand side in equation (\ref{eq:scalartensormodes}) is massively amplified, while the momentum perturbations are strongly suppressed. 
Thus we have found an example of a model in which the spacetime is rendered smooth via accelerated expansion, but where the background solution is (to a good approximation) not affected by the perturbations, thus also without the possibility for the run-away behaviour of eternal inflation. Note that eternal inflation is thought to happen because rare, but large quantum fluctuations change the background evolution by prolonging the smoothing phase in certain regions, with these regions becoming dominant due to the high expansion rate (we should bear in mind though that this argument is based on extrapolating linearised perturbation theory to the limit of its range of validity). In the absence of these large classicalized fluctuations, the background evolution will be essentially unaffected and will proceed as in the purely classical theory. This property certainly deserves further consideration in the future. Note also that for our specific model the large $\epsilon$ limit corresponds to the limit where the scalar field is conformally coupled to gravity ($\xi \approx \frac{1}{6}$), see Eqs. \eqref{eq:F} and \eqref{eq:epsilon_xi}.

When $\epsilon$ is smaller, a certain amount of squeezing will occur -- in particular, although the mode functions themselves become small as $|k\tau| \rightarrow 0,$ the spread in the momenta is enlarged\footnote{We thank an anonymous referee for pointing out this interesting feature.}. This squeezing is of a different type than the familiar one in inflation (where the field value is enlarged, and the spread in momenta suppressed), and an interesting question will be to determine to what extent such fluctuations become classical (note that in contrast to ordinary inflation, where it grows enormously \cite{Polarski:1995jg}, here the product $|v_J| |\pi_v|$ tends to a small constant at late times and thus the uncertainty remains near the quantum minimum), and to what extent they may backreact on the background evolution. We leave these questions for future work. Certainly, for sufficiently large $\epsilon,$ the adiabatic field will not contribute significantly, and we must introduce an additional ingredient in order to generate nearly scale-invariant density perturbations.

\subsection{Non-minimal entropic mechanism in Jordan frame}

In order to obtain a nearly scale-invariant spectrum for the scalar perturbations a second field has to be introduced. There are two possibilities that have been studied extensively in the ekpyrotic literature: either one introduces an unstable direction in the potential \cite{Notari:2002yc,Finelli:2002we,Lehners:2007ac,Koyama:2007mg}, or one allows for a non-minimal kinetic coupling between the two scalars \cite{Li:2013hga,Qiu:2013eoa,Fertig:2013kwa,Ijjas:2014fja,Levy:2015awa}. In both cases nearly scale-invariant entropy perturbations can be generated during the ekpyrotic phase, and these can then be converted to adiabatic curvature perturbations subsequently. Here we will discuss the case of non-minimal coupling, and we will show that it carries over into the context of conflation.

In Einstein frame, one starts with an action of the form \cite{Li:2013hga,Qiu:2013eoa}
\be \label{eq:nonmin}
S=\int d^4x \sqrt{-g} \left[\frac{1}{2}R-\frac{1}{2}g^{\m \n}\p_{\m} \f \p_{\n} \f -
\frac{1}{2}g^{\m \n} e^{-b \f} \p_{\m} \chi \p_{\n} \chi+V_{0}e^{-c\f} \right]\,.
\ee
In the ekpyrotic background, the second scalar $\chi$ is constant. One can then see from the scaling solution \eqref{eq:phi} that when $b=c$ the non-minimal coupling mimics  an exact de Sitter background $e^{-b\phi} \propto 1/t^2$ for the fluctuations $\delta \chi$ (which correspond to gauge-invariant entropy perturbations), which are then amplified and acquire a scale-invariant spectrum. When $b$ and $c$ differ slightly, a small tilt of the spectrum can be generated. 

Transforming the action (\ref{eq:nonmin}) to Jordan frame, we obtain
\be
S_J=\int d^4x \sqrt{-g_J}\left[\x \F^2\frac{R_J}{2}+\frac{1}{2}g_J^{\m \n}\p_{\m} \F \p_{\n} \F -
\frac{1}{2}g_J^{\m \n}  \x^{\frac{\g c -b}{\g c}} \F^{\frac{2\g c -2b}{\g c}}\p_{\m} \chi \p_{\n} \chi
+V_{J,0}\F^{4-\frac{2}{\g}}\right]\,.
\ee
The background equations of motion read
\be
\Box \F+\frac{\g c-b}{\g c}\x^{\frac{\g c -b}{\g c}} \F^{\frac{\g c -2b}{\g c}} g_J^{\m \n}\p_{\m} \chi \p_{\n} \chi-\frac{1}{2}F(\F)_{,\F}R_J  +V(\f)_{J,\F}=0,
\ee
\be
\Box \chi -\frac{2\g c-2b}{\g c} \frac{\F'}{\F} \chi'-2a_J^2 
\xi^{\frac{b-\g c}{\g c}} \F^{\frac{2b-2\g c}{\g c}}  V(\F)_{J,\chi}=0.
\ee
Since the potential is again independent of  $\chi,$ we still have the background solution  $\chi =constant.$ To first order, the equation of motion for the (gauge-invariant) entropy perturbation $\de \chi$ is given by
\be
\de \chi''+ \left(2\frac{a_J'}{a_J}+n \frac{\F'}{\F} \right)\de \chi'+k^2 \de \chi=0\,,
\ee
with $n=\frac{2\g c -2b }{\g c} $. We introduce the canonically normalised variable $v_{Js},$
\be
v_{Js}= a_J \F^{\frac{n}{2}} \de \chi\,,
\ee
whose Fourier modes (dropping the subscript $k$) satisfy the mode equation
\be
v_{Js}''+ \left[k^2+\frac{n}{2} \frac{\F'^2}{\F^2}-\frac{n^2}{4} \frac{\F'^2}{\F^2}+\frac{a_J''}{a_J} (3n \xi-1) -a_J^2\frac{n}{2} \frac{V_{J,\F}}{\F} \right]v_{Js}=0\,.
\ee
Here we have made use of the background equation for $\Phi.$ Plugging in our conflationary background, and using the notation $\Delta=\frac{b}{c}-1$ so that $n=2\frac{\g-\Delta-1}{\g},$ we obtain
\be
v_{Js}''+\left ( k^2-\frac{1}{(\ep-1)^2\tau^2} \left [2-(4+3 \Delta) \ep+(2+3\Delta+\Delta^2)\ep^2 \right ]  \right )v_{Js}=0
\ee
This equation can be solved as usual by $\sqrt{-\tau}$ multiplied by a Hankel function of the first kind with index
\be
\nu=\frac{3}{2}+\frac{\Delta \ep}{\ep-1}\,,
\ee
which translates into a spectral index
\be
n_s -1 = 3 -  2 \nu = - 2 \Delta \frac{\ep}{(\ep -1)}\,.
\ee
The spectrum is independent of $\g$, and in fact it coincides precisely with the spectral index obtained in Einstein frame \cite{Fertig:2013kwa}. Thus, even for this two-field extension, the predictions for perturbations are unchanged by the field redefinition from Einstein to Jordan frame. Note that for models of this type there is no need for an unstable potential, as considered in earlier ekpyrotic models. Also, given that the action does not contain terms in $\chi$ of order higher than quadratic, the ekpyrotic phase does not produce non-Gaussianities. However, the subsequent process of converting the entropy fluctuations into curvature fluctuations (which we assume to occur via a turn in the scalar field trajectory after the end of the conflationary phase) induces a small contribution $|f_{NL}^{local}| \approx 5$ \cite{Lehners:2007wc,Lehners:2008my}, and potentially observable negative $|g_{NL}^{local}| \approx \mathcal{O}(10^2) - \mathcal{O}(10^3)$, as long as the non-minimal field space metric progressively returns to trivial \cite{Fertig:2015ola}, in agreement with observational bounds \cite{Ade:2015cp,Ade:2015ng}. %
It would be interesting to study this and perhaps new conversion mechanisms in more detail from the point of view of the Jordan frame.

\section{Discussion}\label{sec:disscn}

We have introduced the idea of conflation, which corresponds to a phase of accelerated expansion in a scalar-tensor theory of gravity. This new type of cosmology is closely related to anamorphic cosmology, in that it also combines elements from inflation and ekpyrosis -- in fact, our model may be seen as being complementary to anamorphic models. In the conflationary model, the universe is rendered smooth by a phase of accelerated expansion, like in inflation. However, the potential is negative, and adiabatic scalar and tensor fluctuations are not significantly amplified, just as for ekpyrosis.   

Several features deserve more discussion and further study in the future: the first is that, as just mentioned, the conflationary phase described here does not amplify adiabatic fluctuations when $\epsilon$ is large (which is rather easy to achieve as one already has $\epsilon > 3$ by definition) and consequently does not lead to eternal inflation and a multiverse. This remains true in the presence of a second scalar field, which generates cosmological perturbations via an entropic mechanism, since the entropy perturbations that are generated have no impact on the background dynamics. In other words, even a large entropy perturbation is harmless, as it does not cause the conflationary phase to last longer, or proceed at a higher Hubble rate, in that region. This provides a new way of avoiding a multiverse and the associated problems with predictivity, and may be viewed as the most important insight of the present work. The second point is that it would be interesting to study the question of initial conditions required for this type of cosmological model, and contrast it with the requirements for standard, positive potential, inflationary models. A third avenue for further study would be to see how cyclic models in Einstein frame get transformed. Finally, it will be very interesting to see if a conflationary model can arise in supergravity or string theory, with for instance the dilaton playing the role of the scalar field being coupled non-minimally to gravity. Being able to stick to negative potentials while obtaining a background with accelerated expansion opens up new possibilities not considered so far in early universe cosmology.

\acknowledgments

We would like to thank Anna Ijjas and Paul Steinhardt for useful discussions. AF and JLL gratefully acknowledge the support of the European Research Council in the form of the Starting Grant Nr. 256994 entitled ``StringCosmOS''.


\bibliographystyle{apsrev}
\bibliography{jordanframe_bib}

\end{document}